\documentclass[10pt,twocolumn,twoside]{IEEEtran} 
\usepackage{amsmath}
\usepackage{epsfig}
\usepackage{verbatim}
\usepackage{amsfonts}
\usepackage{amsthm}
\newtheorem{thm}{Theorem}

\newtheorem{lemma}{Lemma}
\newtheorem{defn}{Definition}

\usepackage{amssymb}

\def\E{\mathbb{E}}
\def\P{\mathbb{P}}
\def\1{\mathbf{1}}
\def \hyp{\underset{H_0}{\overset{H_1}{\gtrless}}}

\newtheorem{remark}{Remark} 

\begin{document}
\title{Minimax-Optimal Bounds for Detectors \\ Based on Estimated Prior Probabilities}

\author{Jiantao~Jiao*,~Lin~Zhang,~\IEEEmembership{Member,~IEEE} and Robert~D.~Nowak,~\IEEEmembership{Fellow,~IEEE}
\thanks{J. Jiao and L. Zhang are with the Department
of Electronic Engineering, Tsinghua University, Beijing, 100084 China. e-mail: (xajjt1990@gmail.com; linzhang@tsinghua.edu.cn).}
\thanks{R. Nowak is with the Department of Electrical and Computer Engineering,
University of Wisconsin, Madison, WI 53706 USA. e-mail: nowak@ece.wisc.edu.}}

\markboth{Submitted to IEEE Transactions on Information Theory}%
{Jiao \MakeLowercase{\textit{et al.}}: Minimax-Optimal Bounds for Detectors Based on Estimated Prior Probabilities}

\maketitle

\begin{abstract}
  In many signal detection and classification problems, we have knowledge of the distribution under each hypothesis, but not the prior probabilities. This paper is aimed at providing theory to quantify the performance of detection via estimating prior probabilities from either labeled or unlabeled training data.  The error or {\em risk} is considered as a function of the prior probabilities.  We show that the risk function is locally Lipschitz in the vicinity of the true prior probabilities, and the error of detectors based on estimated prior probabilities depends on the behavior of the risk function in this locality.  In general, we show that the error of detectors based on the Maximum Likelihood Estimate (MLE) of the prior probabilities converges to the Bayes error at a rate of $n^{-1/2}$, where $n$ is the number of training data.  If the behavior of the risk function is more favorable, then detectors based on the MLE have errors converging to the corresponding Bayes errors at optimal rates of the form $n^{-(1+\alpha)/2}$, where $\alpha>0$ is a parameter governing the behavior of the risk function with a typical value $\alpha = 1$. The limit $\alpha \rightarrow \infty$ corresponds to a situation where the risk function is flat near the true probabilities, and thus insensitive to small
errors in the MLE; in this case the error of the detector based on the MLE converges to the Bayes error exponentially fast with $n$. We show the bounds are achievable no matter given labeled or unlabeled training data and are minimax-optimal in labeled case.
\end{abstract}

\section{Introduction}
\IEEEPARstart{I}n many signal detection and classification problems the conditional distribution under each hypothesis is known, but the prior probabilities are unknown. For example, we may have a good model for the symptoms of a certain disease, but might not know how prevalent the disease is. There are two ways to proceed: 
\begin{enumerate}
\item Neyman-Pearson detectors
\item Estimate prior probabilities from training data
\end{enumerate}

Neyman-Pearson detectors are designed to control one type of error while minimizing the other.  Detectors based on estimating prior probabilities aim to achieve the performance of the Bayes detector (see, e.g. Devroye, Gyorfi, and Lugosi\cite{Devroye}). We study this second approach and provide theory to quantify the performance of detectors based on estimating prior probabilities from training data. We will focus on simple binary hypotheses and minimum probability of error detection, but the theory and methods can be extended to handle other error criteria that weight different error types and to $m$-ary detection problems. This problem can be viewed as a special case of the classification problem in machine learning in which we have knowledge of the density under each hypothesis.  These conditional densities are called the {\em class-conditional} densities, in the parlance of machine learning, and we will use this terminology here.  Detectors based on ``plugging-in'' the Maximum Likelihood Estimate (MLE) of the prior probabilities are simply a special case of the well-known plug-in approach in statistical learning theory.  We use this connection to develop upper and lower bounds on the performance of detectors based on the MLE of prior probabilities.

Let us first introduce some notations for the problem.  Let $X \in \mathbb{R}^d$ denote a signal and consider a binary hypothesis
testing problem
\begin{eqnarray*}
H_0 & : & X \sim p_0 \\
H_1 & : & X \sim p_1 ,
\end{eqnarray*}
where $p_0$ and $p_1$ are {\em known} probability densities on $\mathbb{R}^d$.  Let $Y$ be a binary random variable indicating which hypothesis $X$ follows, and define $q:=\P(Y=1)$, the probability that hypothesis $H_1$ is true.  The Bayes detector is defined by the likelihood ratio test
$$\frac{p_1(X)}{p_0(X)} \hyp \frac{1-q}{q} \ , $$
and it minimizes the probability of error.

Let $\Lambda(x):=p_1(x)/p_0(x)$ and define the {\em regression function} $\eta(x)$:
\[\eta(x) := P(Y =1|X = x) = \frac{qp_1(x)}{(1-q)p_0(x)+qp_1(x)},\]
then the Bayes detector can be expressed as
\[f^*(x) = \mathbf{1}_{\{\eta(x)\geq 1/2\}}.\]

Note that $\eta(x)$ is parameterized by the prior probability $q$. Let us consider the probability of error, or {\em risk}, as a function of this parameter.  For any feasible prior probability $q' $, let $R(q')$ denote the risk (probability of error) incurred by using $q'$ in place of $q$.  The value $q$ defined above produces the minimum risk.  The difference $R(q')-R(q)$ quantifies the suboptimality of $q'$. The quantity $R(q')$ can be expressed as:
\[R(q') = qP_1(q')+(1-q)P_0(q'),\]
where 
\begin{eqnarray*}
P_1(q') & := & \mathbb{P}(\Lambda(x)< (1-q')/q'|H_1) \\ 
& = & \int \mathbf{1}_{\{\Lambda(x)< (1-q')/q'\}}p_1(x)dx\\
P_0(q')& := & \mathbb{P}(\Lambda(x)\geq (1-q')/q'|H_0) \\ 
& = & \int \mathbf{1}_{\{\Lambda(x)\geq (1-q')/q'\}}p_0(x)dx
\end{eqnarray*}

Assume there is a joint distribution $\pi = \pi_{XY}$ over the signal $X$ and label $Y$. This distribution determines both the class-conditional densities (by conditioning on $Y=0$ or $Y=1$) and the prior probabilities (by marginalizing over $X$).  Suppose we have $n$ training data distributed independent and identically according to $\pi$.  We will consider cases with ``labeled'' $\{(X_i,Y_i)\}_{i=1}^n$ or ``unlabeled'' $\{X_i\}_{i = 1}^n$ data and use them to estimate the {\em unknown} prior probability $q$. Let $\widehat{q}$ stand for the MLE of $q$ based on training data, the risk of the detector based on $\widehat q$ is $R(\widehat q)$.  Note that $R(\widehat q)$ is a random variable and it is greater than or equal to $R(q)$.  The goal of this paper is to bound the difference $\E [R(\widehat q)]-R(q)$,
where $\E$ is the expectation operator, and to provide lower bounds on the performance of any detector derived from knowledge of the class-conditional densities and the training data.
The difference $\E [R(\widehat q)]-R(q)$ is usually called the {\em excess risk} or {\em regret},
and it is a function of $n$. 

Statistical learning theory is typically concerned with the construction estimators based on {\em labeled } training data without prior knowledge of class-conditional densities. There are two common approaches: plug-in rules and empirical risk minimization (ERM) rules (see, e.g., Devroye, Gyorfi, and Lugosi\cite{Devroye} and Vapnik\cite{Vapnik1}). Statistical properties of these two types of classifiers as well as of other related ones have been extensively studied (see Aizerman, Braverman, and Rozonoer\cite{Aizerman}, Vapnik and Chervonenkis\cite{Vapnik}, Vapnik\cite{Vapnik1}\cite{Vapnik2}, Breiman, Friedman, Olshen, and Stone\cite{Breiman}, Devroye, Gyorfi, and Lugosi\cite{Devroye}, Anthony and Bartlett\cite{Anthony}, Cristianini and Shawe-Taylor\cite{Cristianini} and Scholkopf and Smola\cite{Scholkopf} and the references therein). Results concerning the convergence of the excess risk obtained in the literature are of the form
\[\mathbb{E}[R(\widehat{f}_n)]-R(f^*) = O(n^{-\beta})\]
where $\beta>0$ is some exponent, and typically $\beta\leq 1/2$ if $R(f^*)\neq 0$. Here $\widehat{f}_n$ denotes the nonparametric estimator of the classifier, $f^*$ denotes the Bayes classifier. Mammen and Tsybakov\cite{Mammen} first showed that one can attain fast rates, approaching $n^{-1}$, and for further results about the fast rates see Koltchinskii\cite{Koltchinskii2006}, Steinwart and Scovel\cite{Steinwart--Scovel2007}, Tsybakov and van de Geer\cite{Tsybakov--Geer2005}, Massart\cite{Massart} and Catoni\cite{Catoni}. The  behavior of the regression function $\eta$ around the boundary $\partial G^* = \{x: \eta(x) = 1/2\}$ has an important effect on the convergence of the excess risk, which has been discussed earlier under different assumptions by Devroye, Gyrofi, and Lugosi\cite{Devroye} and Horvath and Lugosi\cite{Horvath}. In this paper, we are consider the ``margin assumption'' introduced in Tsybakov\cite{Tsy}. In Audibert and Tsybakov\cite{Tsy2}, they showed there exist plug-in rules converging with {\em super-fast} rates, that is, faster than $n^{-1}$ under the margin assumption in Tsybakov\cite{Tsy}. In our case, which can be viewed as a special case of plug-in rule, we take advantage of Lemma 3.1 in\cite{Tsy2}.

Our main results can be summarized as follows. No matter given labeled or unlabeled data, we show the excess risk converges and deduce the rate of this convergence.  The convergence rate depends on the local behavior of the function $R(\widehat{q})$ near $q$, which is determined by the behavior of $\eta(x)$ in the vicinity of $\eta(x) = 1/2$.  In general, $R$ is locally Lipschitz at $q$, and the convergence rate is proportional to $n^{-1/2}$.  If $R$ is smoother/flatter at $q$, then the convergence rate can be much faster taking the form $n^{-(1+\alpha)/2}$, where $\alpha >0$ is a parameter reflecting the smoothness of $R$ at $q$. The value $\alpha = 1$ is a typical value and we actually have $n^{-1}$ convergence rate under mild conditions. The limit $\alpha \rightarrow \infty$ corresponds to a situation where the risk function is flat near the true probabilities, and thus insensitive to small
errors in the estimate of prior probabilities, in which case the detector based on the MLE converges to the Bayes error exponentially fast with $n$. We also show that the convergence rates are minimax-optimal given labeled data.  Fig.~1 depicts three cases illustrating the smoothness conditions and corresponding $\eta(x)$ considered in the paper.

\begin{figure}[htb]
\begin{minipage}[b]{0.32\linewidth}
  \centering
  \centerline{\epsfig{figure=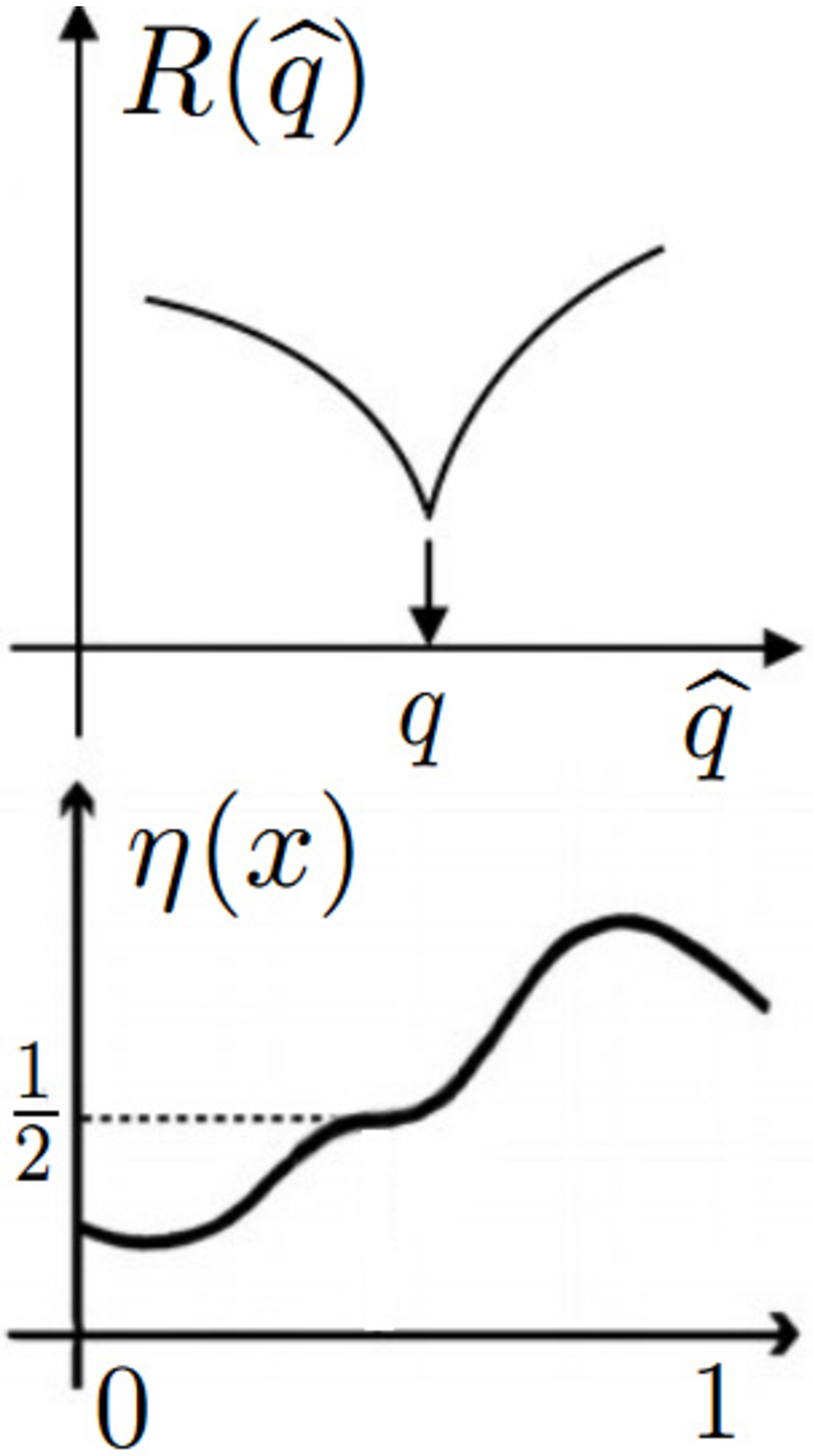,height=5 cm}}
  \centerline{(a) difficult case}\medskip
\end{minipage}
\begin{minipage}[b]{.32\linewidth}
  \centering
  \centerline{\epsfig{figure=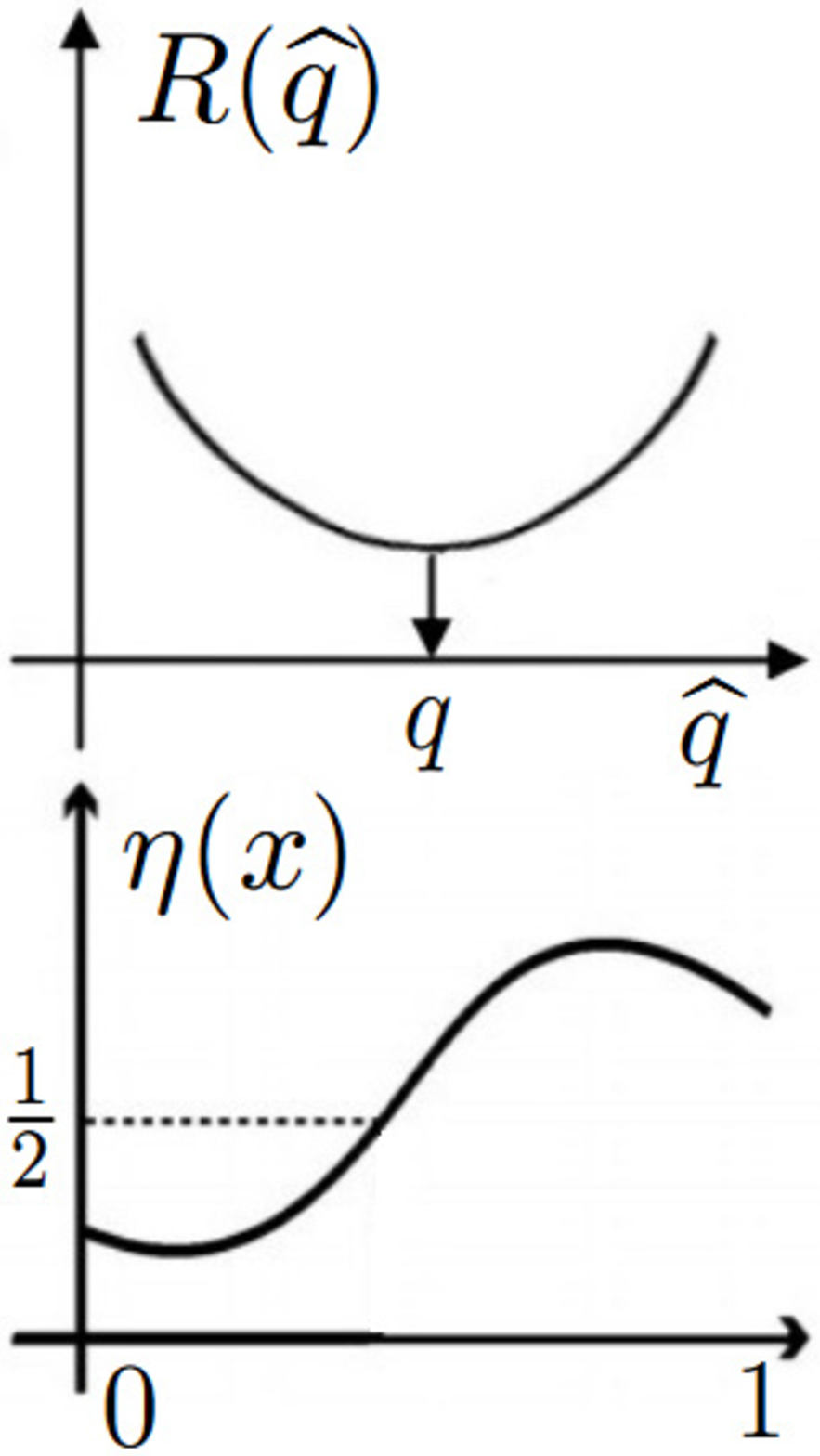,height=5 cm}}
  \centerline{(b) moderate case}\medskip
\end{minipage}
\hfill
\begin{minipage}[b]{0.32\linewidth}
  \centering
  \centerline{\epsfig{figure=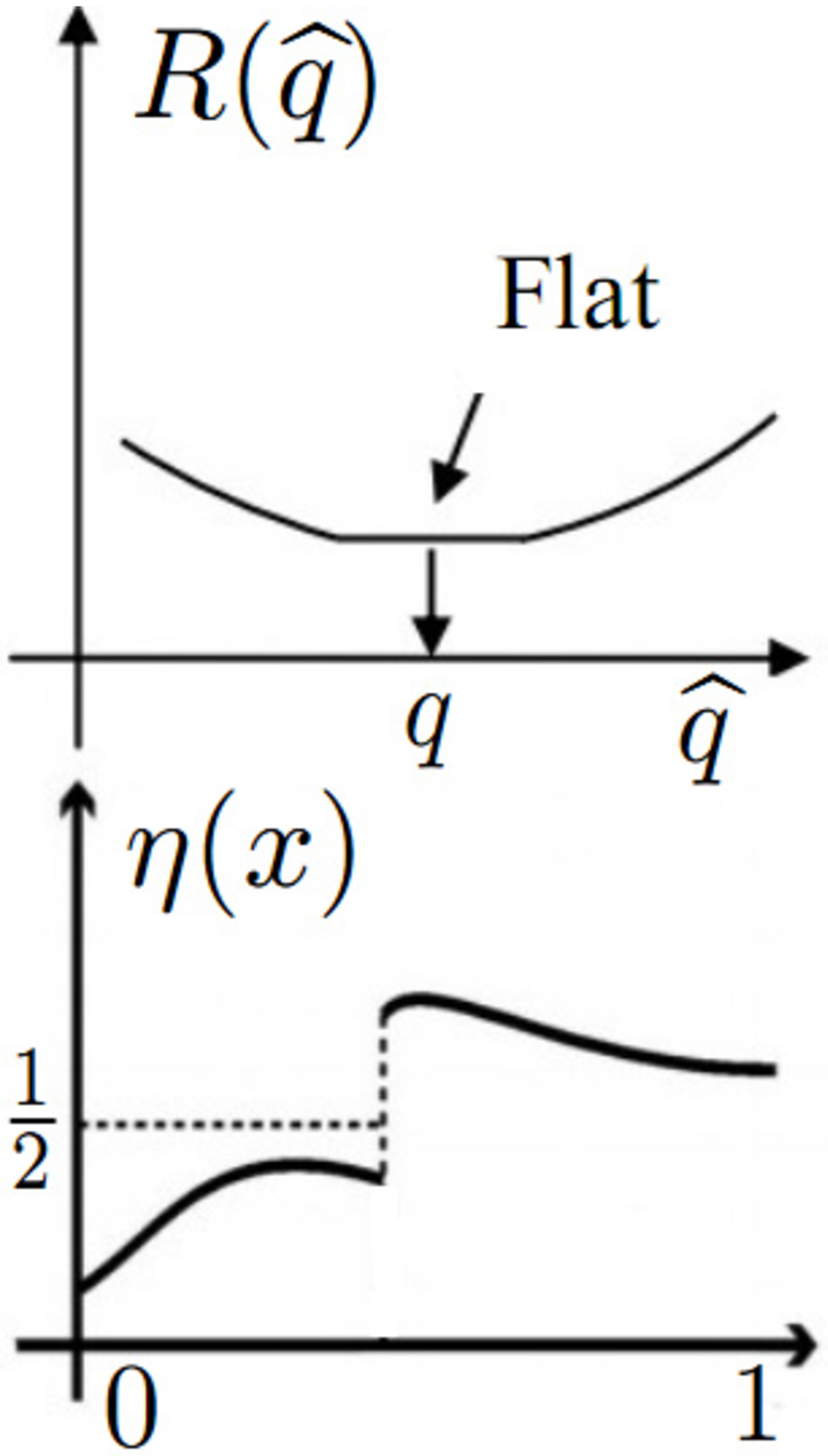,height=5 cm}}
  \centerline{(c) best case}\medskip
\end{minipage}

\centering
\caption{Examples of $R(\widehat{q})$ and corresponding $\eta(x)$ leading to different convergence rates}
\label{fig:res}
\end{figure}

The paper is organized as follows. In Section II and III,we discuss the minimax lower bounds and upper bounds achieved by MLE with labeled data. Section IV discusses convergence rates when we only have unlabeled training data. Section V compares our results with those in standard passive learning and makes final remarks on our work.

\section{Convergence Rates in General Case with Labeled Data} \label{sec.general}
This section discusses the convergence rates of proposed detector trained with labeled data without any assumptions. Let $\widehat{q}$ be the MLE of $q$, i.e.
\[\widehat{q} =  (\sum_{i = 1}^n\mathbf{1}_{\{Y_i = 1\}})/n,\]
define 
\[\mathcal{P}:=\{(p_1,p_0,q)\},\]
where $p_1,p_0$ are class-conditional densities and $q$ is prior probability.

We set up a minimax lower bound:
\subsection{Minimax Lower Bound}
\begin{thm} \label{thm.lower--general}
There exists a constant $c>0$ such that
\[\inf\limits_{\widehat{q}}\sup\limits_{\mathcal{P}}\mathbb{E}[R(\widehat{q})]-R(q)\geq cn^{-1/2},\]
where $\sup\limits_{\mathcal{P}}$ takes supremum over all possible triples $(p_1,p_0,q)$ and $\inf\limits_{\widehat{q}}$ denotes the infimum over all possible estimators of $q$ derived from $n$ samples of training data with the prior knowledge of class-conditional densities.
\end{thm}

Theorem~\ref{thm.lower--general} can be viewed as a corollary of Theorem~\ref{thm.lower--faster} (given in the following section) if we take $\alpha = 0$ and remove constraints on $p_1(x),p_0(x)$ and $q$ in Theorem~\ref{thm.lower--faster}.

\subsection{Upper Bound}
\begin{thm}\label{thm.general--upperbound}
If $\widehat{q}$ is MLE of $q$, we have
\[\sup\limits_{\mathcal{P}}\mathbb{E}[R(\widehat{q})]-R(q)\leq \frac{1}{2}n^{-1/2}.\]
\end{thm}

\begin{IEEEproof}
Define {\em parametrized} risk function as
\[\widehat{R}(q_1;q_2):=q_2P_1(q_1)+(1-q_2)P_0(q_1),\]
following the proof showing $q = \arg\min\limits_{\widehat{q}}R(\widehat{q})$, we know
\[\widehat{q} = \arg\min\limits_{q'} \widehat{R}(q';\widehat{q}).\]

We express the excess risk as
\begin{eqnarray*}
\mathbb{E}[R(\widehat{q})-R(q)] & = & \mathbb{E}[\widehat{R}(\widehat{q};q)-\widehat{R}(q;\widehat{q})]\\
& \leq & \mathbb{E}[\widehat{R}(\widehat{q};q)-\widehat{R}(\widehat{q};\widehat{q})],
\end{eqnarray*}
if we write $\widehat{R}(\widehat{q};q)-\widehat{R}(\widehat{q};\widehat{q})$ explicitly as follows
\[\widehat{R}(\widehat{q};q)-\widehat{R}(\widehat{q};\widehat{q}) = (q-\widehat{q})(P_1(\widehat{q})-P_0(\widehat{q})),\]
thus we have
\begin{eqnarray*}
\mathbb{E}[R(\widehat{q})-R(q)] & \leq & \mathbb{E}[(q-\widehat{q})(P_1(\widehat{q})-P_0(\widehat{q}))] \\
& \leq & \mathbb{E}[|q-\widehat{q}|]\\
& \leq & \sqrt{\mathbb{E}[(q-\widehat{q})^2]}\\
& = & \sqrt{\frac{q(1-q)}{n}}\\
& \leq & \frac{1}{2}n^{-1/2},
\end{eqnarray*}
which completes the proof of Theorem~\ref{thm.general--upperbound}.
\end{IEEEproof}

\begin{remark}
General results in this section also apply when $p_i(x),i = 0,1$ are probability mass functions (pmf). In this case, we can write $p_i(x),i = 0,1$ as summation of a series of weighted {\em Dirac Delta functions}, i.e., \[p_i(x) = \sum\limits_j w_{i,j}\delta(x - x_j),\]
then all of the arguments above hold.
\end{remark}

\section{Faster Convergence Rates with Labeled Data} \label{sec.faster}
In Section~\ref{sec.faster} and Section~\ref{sec.unlabeled}, without loss of generality, we assume the true prior probability $q$ lies in closed interval $[\theta,1-\theta]$, where $\theta$ is an arbitrarily small positive real number. The reason why we need this assumption is explained in Section~\ref{subsec.polynomial}.

Define the {\em trimmed} MLE of $q$ as
\[\widehat{q} := \arg\max\limits_{q\in [\theta,1-\theta]} q^{\sum_{i = 1}^n Y_i}(1-q)^{\sum_{i = 1}^n (1-Y_i)},\]
and construct the regression function estimator $\widehat{\eta}_n(x;\widehat{q})$ as
\[\widehat{\eta}_n(x;\widehat{q}) = \frac{\widehat{q}p_1(x)}{(1-\widehat{q})p_0(x)+\widehat{q}p_1(x)}.\]

The accuracy of $\widehat{\eta}_n(x;\widehat{q})$ is closely related to that of estimating $q$ from $n$ training data. We set up a lemma to describe the Lipschitz property of $\widehat{\eta}_n(x;\widehat{q})$ as a function of $\widehat{q}$.
\begin{lemma} \label{lemma.lipschitz}
The regression function estimator $\widehat{\eta}_n(x)$ satisfies Lipschitz property as a function of $\widehat{q}$
\[\forall q_1,q_2 \in [\theta,1-\theta], \sup\limits_{x\in\mathbb{R}^d}|\widehat{\eta}_n(x;q_1)-\widehat{\eta}_n(x;q_2)|\leq L|q_1-q_2|,\]
where $L = 1/(4\theta(1-\theta))$.
\end{lemma}
\begin{IEEEproof}
Denote $f(t,x) = tp_1(x)/(tp_1(x)+(1-t)p_0(x))$, we are interested in the partial derivative of $f$ over $t$:
\[\frac{\partial f}{\partial t} = \frac{p_0p_1}{(tp_1(x)+(1-t)p_0(x))^2}\geq 0.\]
Since $t\in [\theta,1-\theta]$, we have
\begin{eqnarray*}
\frac{p_0p_1}{(tp_1(x)+(1-t)p_0(x))^2} & \leq & \frac{p_0p_1}{(2\sqrt{t(1-t)p_1p_0})^2} \\
& \leq & \frac{1}{4\theta(1-\theta)},
\end{eqnarray*}
thus
\[\forall q_1,q_2\in [\theta,1-\theta], \sup\limits_{x\in\mathbb{R}^d}|\widehat{\eta}_n(x;q_1)-\widehat{\eta}_n(x;q_2)|\leq L|q_1-q_2|.\]
where $L = 1/(4\theta(1-\theta))\geq 1$.
\end{IEEEproof}

\begin{remark}
On the decision boundary, we have
\[qp_1(x) = (1-q)p_0(x),\]
which makes the inequality shown in the proof of Lemma~\ref{lemma.lipschitz} hold equality, thus we know the Lipschitz constant $L$ cannot be further improved.
\end{remark}

\subsection{Polynomial Rates} \label{subsec.polynomial}
Tsybakov\cite{Tsy} introduced a parametrized margin assumption denoted as Assumption (MA):

There exist constants $C_0>0$, $c>0$, and $\alpha\geq 0$, such that when $\alpha<\infty$, we have
\[
P_X(0<|\eta(X)-\frac{1}{2}|\leq t)\leq C_0t^{\alpha}\quad \forall t>0,
\]
when $\alpha = \infty$, we have
\[
P_X(0<|\eta(X)-\frac{1}{2}|\leq c) = 0
\]

Denote 
\[\mathcal{P}_{\theta,\alpha}: = \{(p_1,p_0,q):\textrm{Assumption (MA) satisfied}\]
\[\textrm{with parameter }\alpha\textrm{ and } q\in [\theta,1-\theta]\},\]
the case $\alpha = 0$ is trivial (no margin assumption) and it is the case explored in Section~\ref{sec.general}. If $d = 1$ and the decision boundary reduces, for example, to one point $x_0$, Assumption (MA) may be interpreted as
\[\eta(x)-\frac{1}{2}\sim (x-x_0)^{1/\alpha}\]
for $x$ close to $x_0$. This interpretation shed light on one fact that $\alpha = 1$ is typical. If $\eta(x)$ is differentiable with non-zero first-order derivative at $x = x_0$, then we know the first-order approximation of $\eta(x)$ in the neighbourhood exists, which means $\alpha = 1$ in this case. When $\eta(x)$ is smoother, for example, if the first-order derivative vanishes at $x = x_0$ but the second-order derivative doesn't, then we have $\alpha = 1/2$. When $\eta(x)$ is not differentiable at $x = x_0$, then we may have $\alpha>1$, for example, when $\alpha = 2$, the derivative of $\eta(x)$ at $x = x_0$ goes to infinity. $\pi_{XY}$ satisfying Assumption (MA) with larger $\alpha$ all have more drastic changes near the boundary $\eta(x) = 1/2$, which makes $R(\widehat{q})$ less sensitive to small errors, leading to faster rates. The $R(\widehat{q})$ and corresponding $\eta(x)$ with typical $\alpha = 1$ in Assumption (MA) are shown in Fig.~1(b). 

We explain the reason why we need to bound the domain of $q$ by showing what determines $C_0$ in Assumption (MA). 

Consider the typical case when $\alpha = 1, d = 1$, calculate the derivative of $\eta(x)$ against $x$ at point $x = x_0$ that the decision boundary reduces to, we have
\[\eta'(x_0) = q(1-q)\frac{p_1'(x_0)p_0(x_0)-p_0'(x_0)p_1(x_0)}{(qp_1(x_0)+(1-q)p_0(x_0))^2}\propto q(1-q).\]

Without loss of generality, suppose the marginal distribution of $X$ is uniform, as the first-order approximation of $\eta(x)$ is
\[\Delta \eta(x)\approx \eta'(x_0)\Delta x, \]
we know
\[P_X(0<|\eta(X)-\frac{1}{2}|\leq t) \propto \frac{1}{\eta'(x_0)}t\propto \frac{t}{q(1-q)}.\]

Then we can see if $q$ goes to zero or one, the constant $C_0$ will approach infinity, which illustrates why we assume $q\in [\theta,1-\theta],\theta>0$ in the beginning of Section~\ref{sec.faster}. Assumption (MA) provides a useful characterization of the behavior of the regression function $\eta(x)$ in the vicinity of the level $\eta(x)$ = 1/2, which turns out to be crucial in determining convergence rates.

First we state a minimax lower bound under Assumption (MA) as follows:

\begin{thm} \label{thm.lower--faster}
There exists a constant $c>0$ such that
\[\inf\limits_{\widehat{q}}\sup\limits_{\mathcal{P}_{\theta,\alpha}}\mathbb{E}[R(\widehat{q})]-R(q)\geq cn^{-(1+\alpha)/2} \]
\end{thm}

The proof is given in Appendix A. It follows the general minimax analysis strategy but is a non-trivial result.

Next we show $n^{-(1+\alpha)/2}$ is also an upper bound. Introduce Lemma 3.1 in Audibert and Tsybakov\cite{Tsy2} which is rephrased as follows:

\begin{lemma}\label{lemma.plug-in}
Let $\widehat{\eta}_n$ be an estimator of the regression function $\eta$ and $\mathcal{P}$ a set of $\pi_{XY}$ satisfying Margin Assumption (MA). If we have some constants $C_1>0$, $C_2>0$, for some positive sequence $a_n$, for $n\geq 1$, any $\delta>0$, and for almost all $x$ w.r.t. $P_X$,
\[\sup\limits_{P\in\mathcal{P}}P(|\widehat{\eta}_n(x)-\eta(x)|\geq \delta)\leq C_1e^{-C_2a_n\delta^2}\]

Then the plug-in detector $\widehat{f}_n = \mathbf{1}_{\{\widehat{\eta}_n\geq 1/2\}}$ satisfies the following inequality:
\[\sup\limits_{P\in \mathcal{P}}\mathbb{E}[R(\widehat{f}_n)]-R(f^{*})\leq Ca_n^{-(1+\alpha)/2}\]

for $n\geq 1$ with some constant $C>0$ depending only on $\alpha,C_0,C_1$ and $C_2$, where $f^{*}$ denotes the Bayes detector.
\end{lemma}

\begin{remark}
Following the proof of Lemma~\ref{lemma.plug-in}, we know $C$ increases as the increase of $C_1$, the increase of constant $C_0$ in Assumption (MA) $C_0$, and the decrease of constant $C_2$.

\end{remark}

\begin{thm} \label{thm.upperbound}
If $\widehat{q}$ is the trimmed MLE of $q$, there exists a constant $C>0$ such that
\[\sup\limits_{\mathcal{P}_{\theta,\alpha}}\mathbb{E}[R(\widehat{q})]-R(q)\leq Cn^{-(1+\alpha)/2} \]
\end{thm}

\begin{IEEEproof}
According to Lemma~\ref{lemma.lipschitz}, we have
\[\sup\limits_{x\in\mathbb{R}^d,\widehat{q},q\in [\theta,1-\theta]}|\widehat{\eta}_n(x;\widehat{q})-\widehat{\eta}_n(x;q)|\leq L|\widehat{q}-q|\]

Combining with Hoeffding's inequality, we have
\[\sup\limits_{\mathcal{P}_{\theta,\alpha}}P(|\widehat{\eta}_n(x)-\eta(x)| \geq \delta)\]
\[\leq \sup\limits_{\mathcal{P}_{\theta,\alpha}}P(|\widehat{q}-q|\geq \frac{\delta}{L}) \leq 2e^{-\frac{2}{L^2}n\delta^2},\]
where $L>0$ is the constant in Lemma~\ref{lemma.lipschitz}. The inequality above shows we can take $C_1 = 2$, $C_2 = 2/L^2$, $a_n = n$ in Lemma~\ref{lemma.plug-in}. According to Lemma~\ref{lemma.plug-in}, we know
\[\sup\limits_{\mathcal{P}_{\theta,\alpha}}\mathbb{E}[R(\widehat{q})]-R(q)\leq Cn^{-(1+\alpha)/2}.\]
\end{IEEEproof}

\begin{remark}
Consider the typical case when $\alpha = 1$. The optimal rate here is $n^{-1}$, which is faster than naive worst case $n^{-1/2}$ shown in Section~\ref{sec.general} and the optimal rate in standard passive learning, $n^{-2/(2+\rho)},\rho>0$ shown in Audibert and Tsybakov\cite{Tsy2}. 
\end{remark}
\begin{remark}
Consider the case when true prior probability $q$ lies near zero or one. This will make the constant $C_0$ in Assumption (MA) go to infinity as shown in the introduction of Assumption (MA), constant $C_2$ go to zero as shown in the proof of Theorem~\ref{thm.upperbound}, which slows down the convergence of excess risk.
\end{remark}

\subsection{Exponential Rates}
We investigate the convergence rates when $\alpha = \infty$ in Assumption (MA). Intuitively as $\alpha$ grows bigger, the rates can be faster than any polynomial rates with fixed degree as is shown in Theorem~\ref{thm.upperbound}.

\begin{thm} \label{thm.exponential--rates}
If $\widehat{q}$ is the trimmed MLE defined above, under Assumption (MA) when $\alpha = \infty$, we have
\[\sup\limits_{\mathcal{P}_{\theta,\infty}}\mathbb{E}[R(\widehat{q})]-R(q)\leq 2e^{-2nc^2/L^2}, \]
where $c$ is the positive constant in Assumption (MA), $L$ is the constant in Lemma~\ref{lemma.lipschitz}.
\end{thm}
\begin{IEEEproof}
According to Lemma~\ref{lemma.lipschitz}, we know as long as $|\widehat{q}-q|\leq c/L, \widehat{q}\in [\theta,1-\theta]$, the error of regression function estimator is bounded uniformly by $c$, incurring no error in detection according to Assumption (MA) when $\alpha =\infty$. The mathematical representation is: $R(\widehat{q}) = R(q),\forall \widehat{q}\in [q-c/L,q+c/L]\cap [\theta,1-\theta]$.

Then we write the excess risk as follows:
\[(\int_{|\widehat{q}-q|\geq \delta}+\int_{|\widehat{q}-q|\leq \delta})[R(\widehat{q})-R(q)]dP\]

where $P$ is the probability measure on sample space $\Omega$ of $\{(X_i,Y_i)\}_{i = 1}^n$.

Taking $\delta = c/L$, the second term vanishes. 

Applying Chernoff's bound, the first term is bounded by $2e^{-2n\delta^2}$, so we conclude
\[\sup\limits_{\mathcal{P}_{\theta,\infty}}\mathbb{E}[R(\widehat{q})]-R(q)\leq 2e^{-2nc^2/L^2}. \]
\end{IEEEproof}

\begin{remark}
When $p_i(x),i = 0,1$ are probability mass functions, if $x$ takes value in $\mathcal{X}$ with $\#\{\mathcal{X}\}< \infty$, then  
\[\inf\limits_{x_i\in\mathcal{X},\eta(x_i)\neq 1/2} |\eta(x_i)-1/2|\geq c>0\]
which means there exists a constant $c>0$ such that 
\[P(0<|\eta(X)-1/2|\leq c) = 0.\]
Based on discussions above, an exponential convergence rate is always guaranteed when $x$ lies in discrete finite domain. If $\#\{\mathcal{X}\}$ is infinite, then we may have finite $\alpha>0$ with optimal convergence rates $n^{-(1+\alpha)/2}$. However, finite $\#\{\mathcal{X}\}$ is the case that often arise in practice.
\end{remark}

\section{Convergence Rates with Unlabeled Data} \label{sec.unlabeled}
In this section, we discuss convergence rates when we only have {\em unlabeled} training data. Relatively speaking, unlabeled data is more likely and easier to be obtained in practice than the labeled, thus convergence rates analysis in this case deserves more attention. Meanwhile, it also helps revealing how much information is stored in $\{X_i\}_{i = 1}^n$ in the training data pairs $\{(X_i,Y_i)\}_{i = 1}^n$. 

In this case, we are faced with a classical parameter estimation problem. Given 
\[X_1,\ldots,X_n \stackrel{iid}{\sim} qp_1(x)+(1-q)p_0(x),\]
we want to construct estimator $\widehat{q}$ to estimate $q$ as efficiently as possible. Here we use the MLE and derive upper bounds under Assumption (MA). 

Before starting the proof, we introduce a standard quantity measuring distances between probability measures. 

\begin{defn}
The {\bf total variation distance} between two probability density functions $p,q$ is defined as follows:
\[V(p,q) = \sup\limits_{A}|\int_{A}(p-q)d\nu| = 1-\int \min(p,q)d\nu\]
where $\nu$ denotes Lebesgue measure on signal space $\mathbb{R}^d$ and $A$ is any subset of the domain.
\end{defn}

We will quantify our results in terms of the total variation distance. Here we assume
\[V(p_1,p_0) \geq V_{\mathrm{min}} >0,\]
ensuring that the two class-conditional densities are not `too' indiscernible, so that it is possible to learn the prior probability $q$ from unlabeled data. For details about how this assumption works please see Appendix B.

Define a class of triples:
\[\mathcal{P}_{\theta,\alpha,V_{\mathrm{min}}}:=\{(p_1,p_0,q):\textrm{Assumption (MA) satisfied}\]
\[\textrm{with parameter }\alpha\textrm{, } q\in [\theta,1-\theta]\textrm{ and }V(p_1,p_0)\geq V_{\mathrm{min}} >0\},\]
and define the trimmed MLE $\widehat{q}$ in this case as
\[\widehat{q}:= \arg\max\limits_{q\in [\theta,1-\theta]} \sum_{i = 1}^n \log (qp_1(x_i) + (1-q)p_0(x_i)).\]

We set up an upper bound for the performance of trimmed MLE $\widehat{q}$:
\begin{thm} \label{thm.upper--unlabeled}
If $\widehat{q}$ is the trimmed MLE defined above, there exists a constant $C>0$ such that
\[\sup\limits_{\mathcal{P}_{\theta,\alpha,V_{\mathrm{min}}}}\mathbb{E}[R(\widehat{q})]-R(q)\leq Cn^{-(1+\alpha)/2}  \]
\end{thm}

The proof of Theorem~\ref{thm.upper--unlabeled} is given in Appendix B. 

\begin{remark}
We can show the calculation of MLE is a convex optimization problem, for which we have efficient methods. 
\end{remark}
\begin{remark}Compared to learning detectors based on labeled data, we need to sacrifice convergence rates by a constant factor when given unlabeled data. Given true prior probability $q$, when $V(p_1,p_0)$ is smaller, the constant $C_2$ in Lemma~\ref{lemma.plug-in} becomes smaller at the same time, which slows down the convergence of excess risk. This phenomenon is discussed in the proof of Theorem~\ref{thm.upper--unlabeled}.
\end{remark}

\section{Final Remarks}
This paper present convergence rates analysis for detectors constructed using known class-conditional densities and estimated prior probabilities using the MLE. All of the bounds are dimension-free. The bounds are minimax-optimal given labeled data and achievable no matter given labeled or unlabeled data. It remains an interesting open question to show the rate $n^{-(\alpha+1)/2}$ is minimax-optimal given unlabeled data under assumption (MA) and the extra assumption on $V(p_1,p_0)$, or to establish the same upper bound on convergence rates for unlabeled case without the extra assumption on $V(p_1,p_0)$ in Section~\ref{sec.unlabeled}. We show the constant factors in convergence rates are mainly influenced by two elements: 

\begin{enumerate}
\item The value of true prior probability
\item Unlabeled data case: $V(p_1,p_0)$
\end{enumerate}

We show a prior probability near zero or one will lead to slower convergence no matter given labeled or unlabeled data, in unlabeled data case, a smaller $V(p_1,p_0)$ leads to slower convergence.

Our results are analogous to those of general classification in statistical learning. Intuitively, learning the class-conditional densities is
the main challenge in standard passive learning and it is sensible for us to say that knowing the class-conditional densities makes the problem relatively easy. The following quantitative results convince us of that. We pick out the fastest-ever rate shown before for standard passive learning under Assumption (MA) in Audibert and Tsybakov\cite{Tsy2} and compare it with our result in table I:

\begin{table}[ht]
\caption{Convergence Rates Comparison under Assumption (MA)} 
\centering 
\begin{tabular}{c | c} 
\hline\hline 
Passive Learning ($p_1,p_0$ unknown) & Passive Learning ($p_1,p_0$ known)\\ [0.5ex] 
\hline 
$n^{-\frac{\alpha+1}{2+\rho}}$  & $n^{-\frac{\alpha+1}{2}}$\\ [1ex] 
\hline 
\end{tabular}
\label{table:nonlin} 
\end{table}

Here $\rho = d/\beta>0$, where $\beta$ is the Holder exponent of $\eta(x)$. The rate $n^{-\frac{\alpha+1}{2+\rho}}$ is obtained with another strong assumption that the marginal distribution of $X$ is bounded from below and above, which isn't necessary here. Here we can see the factor $\rho$ reflects the price we have to pay for not knowing class-conditional densities and it is directly related to the complexity of non-parametrically learning the density functions. 

\section{Acknowledgement}
The authors thank the reviewers for their helpful comments, especially for raising the question of minimax optimality in the unlabeled data case.

\appendices
\section{Proof of Theorem~\ref{thm.lower--faster}}
The proof strategy follows the idea of standard minimax analysis introduced in Tsybakov\cite{Tsy3} and consists in reducing the problem of classification to a hypothesis testing problem. In this case, it suffices to consider two hypotheses. Here, we have to pay extra attention to the design of hypotheses because we have access to class-conditional densities, which puts extra constraint on hypotheses design. We rephrase a bound from Tsybakov\cite{Tsy3}:
\begin{lemma} \label{lemma.tsybakov}
Denote $\mathcal{P}$ the class of joint distributions represented by triples $(p_1,p_0,q)$ where $(p_1,p_0)$ are class-conditional densities and $q$ is prior probability. Associated with each element $(p_1,p_0,q)\in \mathcal{P}$, we have a probability measure $\pi_{XY}$ defined on $\mathbb{R}^d\times \{0,1\}$. Let $d(\cdot,\cdot):\mathcal{P}\times\mathcal{P}\rightarrow \mathbb{R}$ be a semidistance. Let $(p_1,p_0,q_0),(p_1,p_0,q_1)\in \mathcal{P}$ be such that $d((p_1,p_0,q_0),(p_1,p_0,q_1))\geq 2a$, with $a>0$. Assume also that KL$(\pi_{XY}(p_1,p_0,q_1)\rVert \pi_{XY}(p_1,p_0,q_0))\leq \gamma$, where $KL$ denotes to the Kullback-Leibler divergence. The following bound holds:
\begin{eqnarray*}
&\quad &\inf\limits_{\widehat{q}}\sup\limits_{\mathcal{P}}P_{\pi_{XY}(p_1,p_0,q)}(d((p_1,p_0,\widehat{q}),(p_1,p_0,q))\geq a)\\
& \geq  & \inf\limits_{\widehat{q}}\max\limits_{j\in\{0,1\}}P_{\pi_{XY}(p_1,p_0,q_j)}(d((p_1,p_0,\widehat{q}),(p_1,p_0,q_j))\geq a)\\
& \geq  & \max(\frac{1}{4}\exp(-\gamma),\frac{1-\sqrt{\gamma/2}}{2})
\end{eqnarray*}
where the infimum is taken with respect to the collection of all
possible estimators of $q$ (based on a sample from $\pi_{XY}(p_1,p_0,q)$ with known class-conditional densities).
\end{lemma}

\begin{figure}[htb]
\begin{minipage}[b]{1.0\linewidth}
  \centering
  \centerline{\epsfig{figure=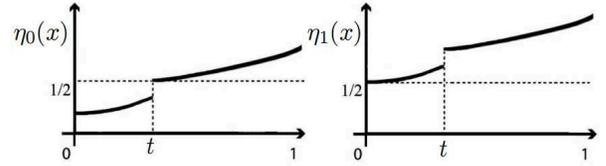,width=8.0cm}}
\end{minipage}
\centering
\caption{Two $\eta(x)$ used for the proof of Theorem~\ref{thm.lower--faster} when $d = 1$}
\label{fig:res}
\end{figure}

Denote $\widehat{G}_n:=\{x:\widehat{\eta}_n(x;\widehat{q})\geq 1/2\}$ where $\widehat{\eta}_n(x;\widehat{q})$ is defined in Section~\ref{sec.unlabeled} and the optimal decision regions as $G_j^*:=\{x:\eta_j(x)\geq 1/2\}$, where the subscript $j$ indicates that the excess risk is being measured with respect to the distribution $\pi_{XY}((p_1,p_0,q_{j})),j = 0,1$. 
Take $\mathcal{P} = \mathcal{P}_{\theta,\alpha}$.  We are interested in controlling the excess risk
\[R_j(\widehat{q})-R_j(q).\]

To prove the lower bound we will use the following class-conditional densities, which allow us to easily attain any desired margin parameter $\alpha$ in Assumption (MA) by adjusting the parameter $\kappa$ below.
\begin{eqnarray*}
p_1(x) & = & \begin{cases}\frac{(1+2cx_d^{\kappa-1})(1-2t^{\kappa-1})}{1-4c(tx_d)^{\kappa-1}}& x\in [0,1]^{d-1}\times [0,t) \\ 1+2c_1(x_d-t)^{\kappa-1} & x\in [0,1]^{d-1}\times [t,1] \\ 0 & x\in \mathbb{R}^d/[0,1]^d\end{cases}\\
p_0(x) & = & \begin{cases} 2-p_1(x) & x\in [0,1]^d \\ 0 & x\in \mathbb{R}^d/[0,1]^d\end{cases}
\end{eqnarray*}
where $x = (x_1,\ldots,x_d)$, $0<c\ll 1,\kappa>1$ are constants. The quantity $0<t\ll 1$ is a small real number which goes to zero as $n\rightarrow \infty$, and will be determined later. It is easy to verify that in order to make $\int_{\mathbb{R}^d}p_i = 1, i = 0,1$ hold, as $t\rightarrow 0$, the number $c_1$ is of order $O(t^{\kappa})$, which also goes to zero. Assigning prior probabilities to $H_1$ and $H_0$
\begin{eqnarray*}
q_0 & = & \frac{1}{2}\\
q_1 & = & \frac{1}{2}+t^{\kappa-1},
\end{eqnarray*}
obviously the margin distribution of $X$, $P_X^{(0)}$ is uniform on $[0,1]^d$, $P_X^{(1)}$ is approximately uniform on $[0,1]^d$. We can compute the regression functions based on equation
\[\eta_j(x) = \frac{q_jp_1(x)}{q_jp_1(x)+(1-q_j)p_0(x)},\]
and have the explicit expressions of $\eta_j(x),j\in \{0,1\}$ as
\begin{eqnarray*}
\eta_0(x) & = & \begin{cases} \frac{(1/2+cx_d^{\kappa-1})(1-2t^{\kappa-1})}{1-4c(tx_d)^{\kappa-1}}& x\in [0,1]^{d-1}\times [0,t) \\ \frac{1}{2}+c_1(x_d-t)^{\kappa-1} & x\in [0,1]^{d-1}\times [t,1] \\ 0 & x\in \mathbb{R}^d/[0,1]^d\end{cases} \\
\eta_1(x)  & = & \begin{cases}\frac{1}{2}+cx_d^{\kappa-1} & x\in [0,1]^{d-1}\times [0,t) \\ \frac{(1+2t^{\kappa-1})(\frac{1}{2}+c_1(x_d-t)^{\kappa-1})}{1+4t^{\kappa-1} c_1(x_d-t)^{\kappa-1}} & x\in [0,1]^{d-1}\times [t,1] \\ 0 & x\in \mathbb{R}^d/[0,1]^d \end{cases}
\end{eqnarray*}
From above we see $G_0^* = [0,1]^{d-1}\times [t,1]$, $G_1^* = [0,1]^d$. Fig.~2 depicts $\eta_j(x),j\in \{0,1\}$ when $d = 1$.

In order to further analyze designed hypotheses, we show that the parameter $\alpha$ in Assumption (MA) for $\eta_j(x),j = 0,1$ is $\alpha = 1/(\kappa-1)$. Consider the case $j = 0$ (the case $j = 1$ is analogous). 

As $\eta_0((0,\ldots,0,t)) = 1/2-\frac{(1-c)t^{\kappa-1}}{1-4ct^{2\kappa-2}}<1/2-(1-c)t^{\kappa-1} = 1/2-\tau^*$, provided $\tau\leq \tau^*$, we have
\begin{eqnarray*}
P_0(0<|\eta_0(X)-\frac{1}{2}|\leq \tau) & = & P_0(0<x_d-t\leq (\frac{\tau}{c_1})^{1/(\kappa-1)})\\
& = & (\frac{\tau}{c_1})^{1/(\kappa-1)}\\
& = & C_{\eta}\tau^{1/(\kappa-1)},
\end{eqnarray*}
where $C_{\eta}>1$. The second step follows since $P_X^{(0)}$ is uniform on $[0,1]^d$. 

Since the excess risk is not a semidistance, we cannot apply Lemma~\ref{lemma.tsybakov} directly, but we can relate excess risk and the symmetric distance measure, and then use the lemma. First we introduce Proposition 1 in Tsybakov\cite{Tsy} rephrased as follows:
\begin{lemma} \label{lemma.tsybakov1}
Assume that $P(0<|\eta(X)-1/2|\leq \tau)\leq C_{\eta}\tau^{\alpha}$ for some finite $C_{\eta}>0$, $\alpha>0$ and all $0<\tau\leq \tau_{*}$, where $\tau_{*}\leq 1/2$. Then we know there exist $c_{\alpha}>0$,$0<\epsilon_{0}\leq 1$ such that
\[R_j(\widehat{q})-R_j(q)\geq c_{\alpha}d_{\Delta}(\widehat{G}_n,G_P^*)^{1+1/\alpha}\]
for all $\widehat{G}_n$ such that $d_{\Delta}(\widehat{G}_n,G_P^*)\leq \epsilon_0\leq 1$, where $c_{\alpha} = 2C_{\eta}^{-1/\alpha}\alpha(\alpha+1)^{-1-1/\alpha}$, $\epsilon_0 = C_{\eta}(\alpha+1)\tau_{*}^{\alpha}$, $d_{\Delta}(\widehat{G}_n,G_P^*) := \int_{\widehat{G}_n\Delta G_P^*}d\mathbf{x}$ is the symmetric distance measure.
\end{lemma}
When $j = 0$, plug in $\tau_{*} = (1-c)t^{\kappa-1}$, since $c$ is very small, we know $\epsilon_{0} = C_{\eta}(1+1/(\kappa-1))(1-c)^{1/(\kappa-1)}t\geq t/2$. Analogously we can show when $j = 1$, $\epsilon_{0}\geq t/2$ also holds.

We now proceed by applying Lemma~\ref{lemma.tsybakov} to the semidistance $d_{\Delta}$ and then use Lemma~\ref{lemma.tsybakov} to control the excess risk. Note that $d_{\Delta}(G_0^*, G_1^*) = t$. Let $P_{0,n}:= P^{(0)}_{X_1,\ldots,X_n;Y_1,\ldots,Y_n}$ be the probability measure of the random variables $\{(X_i,Y_i)\}_{i = 1}^n$ under hypothesis 0 and define analogously $P_{1,n}:= P^{(1)}_{X_1,\ldots,X_n;Y_1,\ldots,Y_n}$. Consider the KL-divergence KL$(P_{1,n}\rVert P_{0,n})$:
\begin{eqnarray*}
\mathrm{KL}(P_{1,n}\rVert P_{0,n}) & = & \mathbb{E}_1[\log \frac{\Pi_{i = 1}^n p_{X_i,Y_i}^{(1)}(X_i,Y_i)}{\Pi_{i = 1}^n p_{X_i,Y_i}^{(0)}(X_i,Y_i)}]\\ 
 & = & \sum_{i = 1}^n \mathbb{E}_1[\log \frac{p_{X_i,Y_i}^{(1)}(X_i,Y_i)}{p_{X_i,Y_i}^{(0)}(X_i,Y_i)}]\\
& = & n\mathbb{E}_1[\log \frac{p_{X,Y}^{(1)}(X,Y)}{p_{X,Y}^{(0)}(X,Y)}],
\end{eqnarray*}
where $\mathbb{E}_1[\log \frac{p_{X,Y}^{(1)}(X,Y)}{p_{X,Y}^{(0)}(X,Y)}]$ can be simplified as
\[\int_{\mathbb{R}^d}q_1p_1(x)\log \frac{q_1 p_1(x)}{q_0 p_1(x)} + \int_{\mathbb{R}^d}(1-q_1)p_0(x)\log \frac{(1-q_1)p_0(x)}{(1-q_0)p_0(x)}\]
\[ = q_1\log \frac{q_1}{q_0}+(1-q_1)\frac{1-q_1}{1-q_0}.\]
The expression in the last line is the KL-divergence between two Bernoulli random variables. It can be easily verified that the KL-divergence between two Bernoulli random variables is bounded as in the following lemma:
\begin{lemma} \label{lemma.bernoulli}
Let $P$ and $Q$ be Bernoulli random variables with parameters, respectively, 1/2-$p$ and 1/2-$q$. Let $|p|,|q|\leq 1/4$, then KL$(P\rVert Q)\leq 8(p-q)^2$.
\end{lemma}

Thus we know
\begin{eqnarray*}
\mathrm{KL}(P_{1,n}\rVert P_{0,n}) & \leq & 8n(t^{\kappa-1})^2\\
& = & 8nt^{2\kappa-2}.
\end{eqnarray*}
Taking $t = n^{-\frac{1}{2\kappa-2}}$, $d((p_1,p_0,\widehat{q}),(p_1,p_0,q_j)): = d_{\Delta}(\widehat{G}_n,G_j^*)$ and using Lemma~\ref{lemma.tsybakov}, we know for $n$ large enough (implying $t$ small),
\begin{eqnarray*}
& & \inf\limits_{\widehat{q}}\max\limits_{j\in\{0,1\}}P_j(d((p_1,p_0,\widehat{q}),(p_1,p_0,q_j))\geq t/2) \\
& \geq & 1/4\exp(-8).
\end{eqnarray*}
Notice in Lemma~\ref{lemma.tsybakov1}, $\epsilon_0\geq t/2$, so we can apply Lemma~\ref{lemma.tsybakov1} to show
\begin{eqnarray*}
& & \inf\limits_{\widehat{q}}\max\limits_{j\in\{0,1\}} P_j(R_j(\widehat{q})-R_j(q)\geq c_{\alpha}(t/2)^{\kappa})\\
& \geq & \inf\limits_{\widehat{q}}\max\limits_{j\in\{0,1\}}P_j(d((p_1,p_0,\widehat{q}),(p_1,p_0,q_j))\geq t/2) \\
& \geq & 1/4\exp(-8).
\end{eqnarray*}
According to Markov's inequality, we conclude
\begin{eqnarray*}
\inf\limits_{\widehat{q}}\sup\limits_{\mathcal{P}_{\theta,\alpha}}\mathbb{E}[R(\widehat{q})-R(q)] & \geq &  c'n^{-\frac{\kappa}{2\kappa-2}} \\
& = &c'n^{-(1+\alpha)/2}
\end{eqnarray*}
where $\alpha = 1/(\kappa-1),c' = \frac{1}{4}e^{-8}c_{\alpha}(\frac{1}{2})^{\frac{\alpha+1}{\alpha}}$.

\section{Proof of Theorem~\ref{thm.upper--unlabeled}}
We introduce two more quantities measuring distances between probability distributions.

\begin{defn}
The {\bf Hellinger distance} between two probability density functions $p,q$ is defined as follows:
\[H(p,q) = (\int(\sqrt{p}-\sqrt{q})^2 d\nu)^{1/2}\]
\end{defn}

\begin{defn}
The {\bf $\chi^2$ divergence} between two probability density functions $p,q$ is defined as follows:
\[\chi^2(p,q) = \int_{pq>0} \frac{p^2}{q}d\nu -1 \]
\end{defn}

As is shown in Tsybakov\cite{Tsy3}, we have the following inequalities
\[V^2(p,q)\leq H^2(p,q)\leq \chi^2(p,q).\]

Define $f(x,q) = qp_1(x)+ (1-q)p_0(x)$, we use Hellinger distance to measure the error of estimating $q$ from training data:
\[r_2(q,q+h):= H(f(x,q),f(x,q+h))\]

We introduce a concentration inequality for MLE, i.e., Theorem I.5.3 in Ibragimov and Has'minskii\cite{Asym} rephrased as follows:
\begin{lemma} \label{lemma.mle}
Let $\mathcal{Q}$ be a bounded interval in $\mathbb{R}$, $f(x,q)$ be a continuous function of $q$ on $\mathcal{Q}$ for $\nu$-almost all $x$ where $\nu$ denotes the Lebesgue measure on $\mathbb{R}^d$, let the following conditions be satisfied:
\begin{enumerate}
\item There exists a number $\xi>1$ such that
\[\sup\limits_{q\in \mathcal{Q}}\sup\limits_{h} |h|^{-\xi}r_2^2(q,q+h) = A <\infty\]
\item For any compact set $K$ there corresponds a positive number $a(K) = a>0$ such that
\[r_2^2(q,q+h)\geq \frac{a|h|^{\xi}}{1+|h|^{\xi}}\quad q\in K,\]
\end{enumerate}
then the maximum likelihood estimator $\widehat{q}$ is defined, consistent and 
\[\sup\limits_{q\in K} P_{q}(|\widehat{q}-q|>\epsilon) \leq B_0e^{-b_0 a n \epsilon^{\xi}},\]
where the positive constants $B_0$ and $b_0$ do not depend on $K$, $n$ is the number of training data.
\end{lemma}

Taking $\mathcal{Q} = K = [\theta,1-\theta]$, it suffices to show the two assumptions in Lemma~\ref{lemma.mle} hold with $\xi = 2$, then we can use Lemma~\ref{lemma.plug-in} to complete the proof.

\begin{IEEEproof}

{\noindent 1. $\sup\limits_{q\in \mathcal{Q}}\sup\limits_{h} |h|^{-2}r_2^2(q,q+h) = A <\infty$}

\begin{eqnarray*}
r_2^2(q,q+h) & \leq & \chi^2(f(x,q+h),f(x,q)) \\
& = & \int f(x,q) + \frac{(p_1-p_0)^2 h^2}{f(x,q)}\\
& &  + 2h(p_1 - p_0)d\nu -1 \\
& = & h^2\int \frac{(p_1 - p_0)^2}{qp_1+ (1-q)p_0}\\
& = & h^2\int_{p_1>p_0} \frac{(p_1 - p_0)^2}{q(p_1 - p_0)+p_0} \\
& & + h^2\int_{p_1<p_0} \frac{(p_0-p_1)^2}{(1-q)(p_0-p_1)+p_1}\\
& \leq & h^2\int_{p_1>p_0} \frac{(p_1-p_0)^2}{q(p_1-p_0)} \\ 
& & + h^2\int_{p_1<p_0} \frac{(p_0-p_1)^2}{(1-q)(p_0 - p_1)}\\
& \leq & h^2(\frac{1}{q} + \frac{1}{1-q})V(p_1,p_0)\\
& = & h^2\frac{1}{q(1-q)}V(p_1,p_0)\\
& \leq & h^2\frac{1}{\theta(1-\theta)}V(p_1,p_0)\\
& \leq & h^2\frac{1}{\theta(1-\theta)}
\end{eqnarray*}

Thus, we verified the first assumption in Lemma~\ref{lemma.mle} by asserting
\[\sup\limits_{q\in [\theta,1-\theta]}\sup\limits_{h}h^{-2}r_2^2(q,q+h) \leq \frac{1}{\theta(1-\theta)} <\infty\]

{\noindent 2. $r_2^2(q,q+h)\geq a|h|^{\xi}/(1+|h|^{\xi})\quad q\in K$}
Since
\begin{eqnarray*}
r_2^2(q,q+h) & \geq & V^2(f(x,q),f(x,q+h))\\
& = & h^2 V^2(p_1,p_0) \\
& \geq &\frac{V^2(p_1,p_0)h^2}{1+h^2}\\
& \geq & \frac{V_{\mathrm{min}}^2h^2}{1+h^2},
\end{eqnarray*}
we can take $a = V_{\mathrm{min}}^2$. Then we can show
\[\sup\limits_{q\in [\theta,1-\theta]} P_{q}(|\widehat{q}-q|>\epsilon) \leq B_0e^{-b_0 a n \epsilon^{2}}\]

Applying Lemma~\ref{lemma.plug-in} by taking $C_1 = B_0$, $C_2 = b_0 a$, $a_n = n$, we complete the proof of Theorem~\ref{thm.upper--unlabeled}. 
\end{IEEEproof}

\begin{remark}
In the proof of Theorem~\ref{thm.upper--unlabeled}, we have
\[1/\left(\int \frac{(p_1-p_0)^2}{qp_1 + (1-q)p_0}\right)\geq \frac{q(1-q)}{V(p_1,p_0)},\]
where the left term is the reciprocal of the fisher information given unlabeled data, and the right term is the fisher information given labeled data divided by $V(p_1,p_0)$. This inequality holds equality when $p_1$ and $p_0$ don't verlap at all. Since the minimum variance of unbiased estimator is described by the reciprocal of fisher information, this inequality shows that the convergence from $\widehat{q}$ to $q$ in unlabeled case can never be faster than that in labeled case, and will be slower if $V(p_1,p_0)$ is small.
\end{remark}

\bibliographystyle{IEEEtran}
\bibliography{stat}
\end{document}